# Enlargement of Memory Window of Si Channel FeFET by Inserting Al$_2$O$_3$ Interlayer on Ferroelectric Hf$_{0.5}$Zr$_{0.5}$O$_2$

Tao Hu, Xiaoqing Sun, Mingkai Bai, Xinpei Jia, Saifei Dai, Tingting Li, Runhao Han, Yajing Ding, Hongyang Fan, Yuanyuan Zhao, Junshuai Chai, Hao Xu, Mengwei Si, Xiaolei Wang, and Wenwu Wang

*Abstract*—In this work, we demonstrate the enlargement of the memory window of Si channel FeFET with ferroelectric Hf$_{0.5}$Zr$_{0.5}$O$_2$ by gate-side dielectric interlayer engineering. By inserting an Al$_2$O$_3$ dielectric interlayer between TiN gate metal and ferroelectric Hf$_{0.5}$Zr$_{0.5}$O$_2$, we achieve a memory window of 3.2 V with endurance of ~10$^5$ cycles and retention over 10 years. The physical origin of memory window enlargement is clarified to be charge trapping at the Al$_2$O$_3$/Hf$_{0.5}$Zr$_{0.5}$O$_2$ interface, which has an opposite charge polarity to the trapped charges at the Hf$_{0.5}$Zr$_{0.5}$O$_2$/SiO$_x$ interface.

## I. Introduction

HAFNIA (HfO$_2$) based silicon channel ferroelectric field-effect transistors (HfO$_2$ Si-FeFET) have received extensive research for non-volatile memories [1-7], thanks to the discovery of ferroelectricity in doped-HfO$_2$ [8]. The memory window (MW) of HfO$_2$ Si-FeFET is mostly about 1-2 V in literature reports [9-12], which does not satisfy its requirements for application in multi-bit memory cells. Recently, a large MW up to 10.5 V was reported in Si-FeFET by optimizing the ferroelectric layer and the gate-side interlayer [13]. However, it did not give the material of the interlayer. And its physical mechanism is still not reported and clarified. To improve the MW, there are generally two kinds of methods. One of the current methods mainly focuses on decreasing the electric field in the bottom SiO$_x$ interlayer between the doped HfO$_2$ ferroelectric and Si channels and consequently suppressing the charge trapping at the doped-HfO$_2$/SiO$_x$ interface [14-17]. The other method focuses on the improvement of the SiO$_x$ quantity. However, a effective method to improve the MW of Si FeFET is still lacking.

In this work, we improve the MW of HfO$_2$ Si-FeFET from about 1 V to over 3 V. We inserts an Al$_2$O$_3$ dielectric layer above the ferroelectric Hf$_{0.5}$Zr$_{0.5}$O$_2$, i.e., the gate stacks are TiN/Al$_2$O$_3$/Hf$_{0.5}$Zr$_{0.5}$O$_2$/SiO$_x$/Si (MIFIS). We also discuss the physical origin.

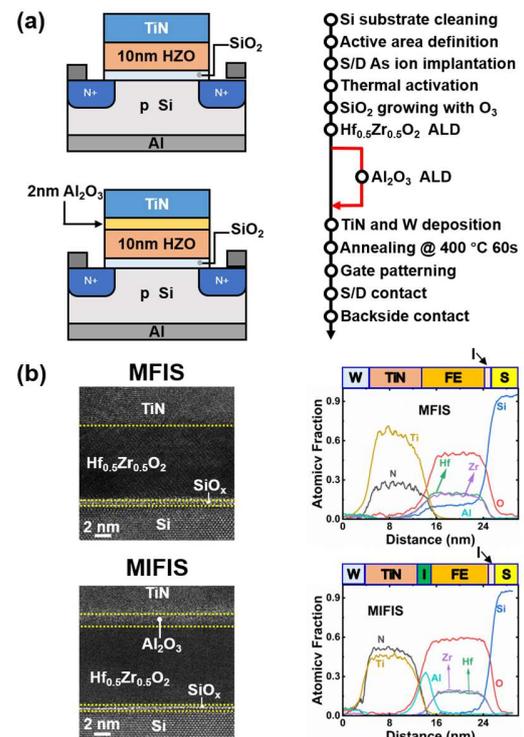

Fig. 1. (a) Schematic of the HfO$_2$ Si-FeFET device structure and fabrication process flow. (b) HRTEM images and EDS of the MIFIS and MFIS structures.

## II. DEVICE FABRICATION AND CHARACTERIZATION

Fig. 1(a) schematically shows the HfO$_2$ Si-FeFET device structure and fabrication process flow in this work. There are two different structures for gate stack. One is TiN/Hf$_{0.5}$Zr$_{0.5}$O$_2$/SiO$_x$/Si (MFIS) as the control sample. The other is TiN/Al$_2$O$_3$/Hf$_{0.5}$Zr$_{0.5}$O$_2$/SiO$_x$/Si (MIFIS).

This work was supported in part by National Key Research and Development Program of China under Grant No. 2022YFB4400300 and in part by the National Natural Science Foundation of China under Grant No. 92264104. (Corresponding author: Xiaolei Wang)

Tao Hu, Xiaoqing Sun, Mingkai Bai, Xinpei Jia, Saifei Dai, Tingting Li, Runhao Han, Yajing Ding, Hongyang Fan, Yuanyuan Zhao, Junshuai Chai, Hao Xu, Xiaolei Wang, and Wenwu Wang are with Institute of microelectronics of the Chinese Academy of Sciences, Beijing 100029, China. The authors are also with University of Chinese Academy of Sciences, Beijing 100049, China (wangxiaolei@ime.ac.cn).

Mengwei Si is with Department of Electronic Engineering, Shanghai Jiao Tong University,Shanghai 200240, China.

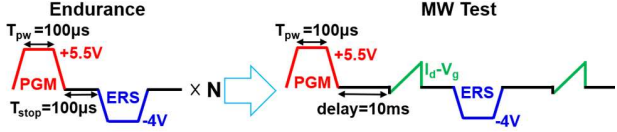

Fig. 2. The pulse sequence of electrical measurement.

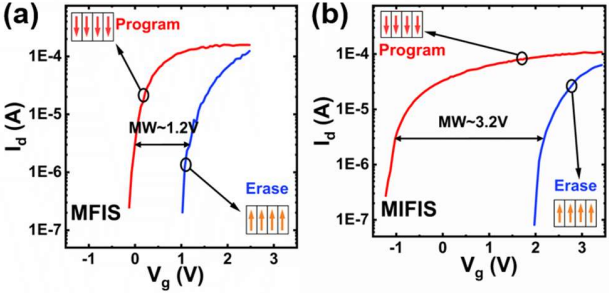

Fig. 3. Measured $I_d$–$V_g$ curves of (a) MFIS and (b) MIFIS devices.

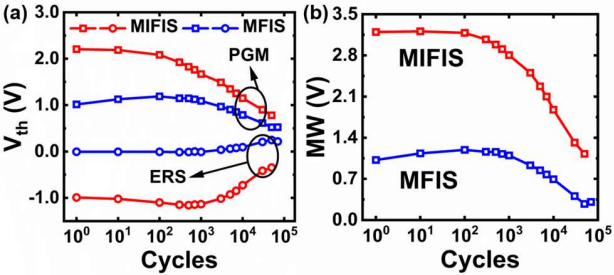

Fig. 4. (a) $V_{th}$ and (b) MW as a function of cycles for the MFIS and MIFIS devices.

The two devices were both fabricated using the gate-last process in an 8-inch p-type silicon wafer. Firstly, B ions with a dose of $8 \times 10^{12}$ cm$^{-2}$ and an energy of 60 keV were implanted into the whole Si substrate to enlarge the doping concentration of the channel. Next, the source and drain regions were defined through photolithography, and then the As ions with a dose of $2 \times 10^{15}$ cm$^{-2}$ and an energy of 50 keV were implanted into the Si substrate to form the source and drain region. After that, the devices were annealed at 1050 °C for 5 s under the N$_2$ atmosphere for dopant activation. Following that was the formation of the gate stacks. After cleaning the substrate with diluted HF, the SiO$_x$ interlayer was grown using O$_3$ oxidation at 300 °C. Then, 10 nm Hf$_{0.5}$Zr$_{0.5}$O$_2$ and 2 nm Al$_2$O$_3$ were grown by using the atomic layer deposition (ALD) at 300 °C. Finally, The 10 nm TiN and 75 nm W were grown by sputtering. Then the devices are annealed at 400 °C in N$_2$ for 60 s to achieve the orthogonal phase. Finally, the forming gas annealing (FGA) was done at 450 °C in 5% - H$_2$/95% - N$_2$. All of the fabrication processes of the MFIS structure were the same as the MIFIS structure, except without the top dielectric interlayer Al$_2$O$_3$. The gate length/width (L/W) of the FeFET in this work is 5/150 μm. The electrical properties were all measured by Keysight B1500A, and the threshold voltage ($V_{th}$) is defined by the constant current method.

Fig. 1(b) shows the High-Resolution Transmission Electron Microscopy (HRTEM) images and the Energy Dispersion Spectrometer (EDS) results of the MIFIS and MFIS structures. For the MIFIS structure, a peak concentration of Al appears at the TiN/Hf$_{0.5}$Zr$_{0.5}$O$_2$ interface, confirming the presence of the Al$_2$O$_3$ layer.

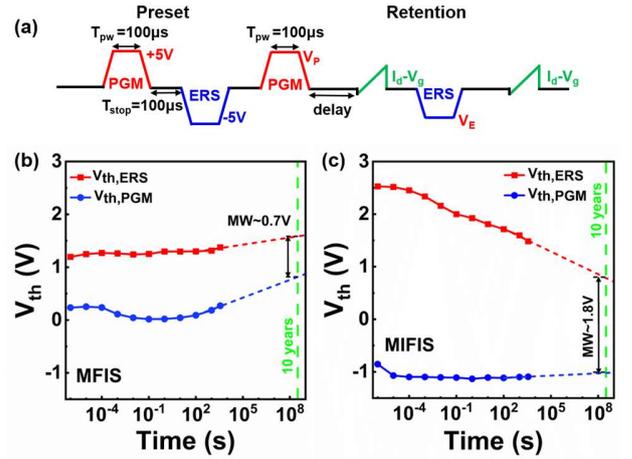

Fig. 5. (a) The measurement waveform of retention. Retention characteristics measured at room temperature of (b) the MFIS structure and (c) the MIFIS structure.

### III. RESULTS AND DISCUSSIONS

Fig. 2 shows the measurement waveform. We use the pulse with the amplitude ($V_a$) of ±5 V and the width ($T_{pw}$) of 100 μs to wake up the FeFET. After that, to study the dependence of the MW on the program/erase voltage amplitude, we change the program voltage from +2 V to +7 V and the erase voltage from -2 V to -5 V with the voltage step being 0.5 V (not shown here). After that, we choose the program and erase voltage amplitudes for the maximum MW for each device. Finally, we determine the program/erase condition as +4/-3.5, 100 μs for the MFIS control sample and +5.5/-4, 100 μs for the MIFIS sample.

Fig. 3 shows the $I_d$-$V_g$ curve of the two devices after waking up. For the control MFIS sample, the MW is 1.2 V. For the MIFIS sample, the MW is 3.2 V. Moreover, the $V_{th}$ after PGM ($V_{th,PGM}$) of the MIFIS device negatively shifts compared to the control MFIS sample, and the $V_{th}$ after ERS ($V_{th,ERS}$) of the MIFIS device positively shifts compared to the control MFIS sample. Thus the insertion of Al$_2$O$_3$ at the TiN/ Hf$_{0.5}$Zr$_{0.5}$O$_2$ interface significantly increases the MW. Fig. 4 shows the endurance characteristics. Both devices breakdown after ~7 × 10$^4$ cycles. The MIFIS sample does not nearly degrade after 10$^3$ cycles. Fig. 5 shows the retention characteristics. The results show that both devices have a retention lifetime beyond 10 years. In addition, the MIFIS device shows relatively large MW degradation after 10 years. Moreover, it is worth noting that the $V_{th}$ after ERS ($V_{th,ERS}$) decreases significantly for the MIFIS device. In summary, the MIFIS device shows significant improvement in the MW and comparable endurance and retention characteristics compared with the control MFIS sample. The above results confirm the effectiveness of the proposed method.

We discuss the physical origin of memory window enlargement by inserting the Al$_2$O$_3$ top interlayer. The origin is charge trapping between the TiN metal gate and the Al$_2$O$_3$/Hf$_{0.5}$Zr$_{0.5}$O$_2$ interface. Fig. 6(a) shows the band digram during the program pulse. The electron trapping occurs between the Si substrate and the $Hf_{0.5}Zr_{0.5}O_2/SiO_x$ interface, as well understood. In addition, the electron tunneling between the TiN metal gate and the $Al_2O_3/Hf_{0.5}Zr_{0.5}O_2$ interface also appears, leaving positive charges at the $Al_2O_3/Hf_{0.5}Zr_{0.5}O_2$ interface. After the program, i.e., the gate voltage returns to 0 V, as shown in Fig. 6(b), positive charges at the $Al_2O_3/Hf_{0.5}Zr_{0.5}O_2$ interface are remained because the $Al_2O_3$ acts as a barrier layer. The positive charges shift the $V_{th}$ negatively. Similarly, Figs. 6(c) and (d) show the erase case. Negative charges are reserved after the erase pulse and positively shift the $V_{th}$. Finally, the above results lead to a large MW.

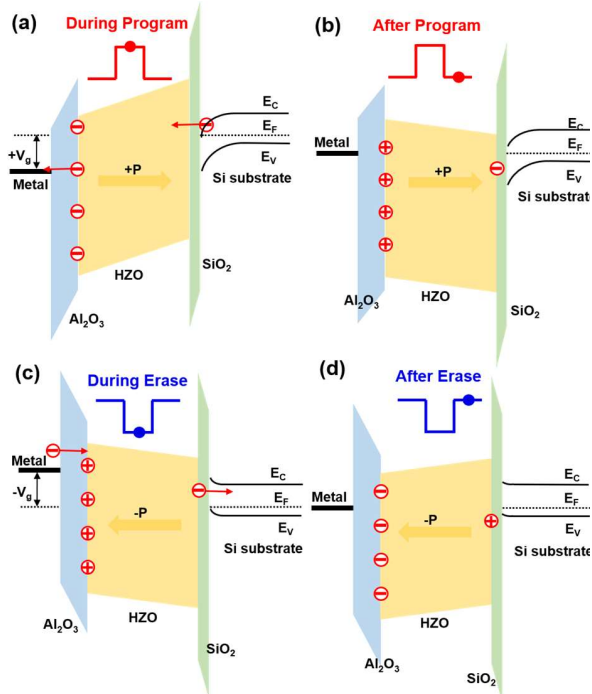

Fig. 6. Band diagram of the MIFIS: (a) during program, (b) after program, (c) during erase and (d) after erase.

The MW of FeFET is usually limited by its coercive voltage ($V_c$). Fig. 3(b) shows the MW of the MIFIS device is 3.2 V, which is larger than the MW determined by the $2V_c$. Therefore, in the MIFIS structure, the $V_c$ is no longer a limiting factor for the FeFET MW. Our work brings a larger MW at lower operating voltages.

## IV. CONCLUSIONS

We increase the memory window of Si channel FeFET with ferroelectric $Hf_{0.5}Zr_{0.5}O_2$ by inserting an $Al_2O_3$ dielectric interlayer between TiN gate metal and ferroelectric $Hf_{0.5}Zr_{0.5}O_2$. The memory window increases to 3.2 V with an endurance of $10^5$ cycles and retention over 10 years. The physical origin of memory window enlargement is charge trapping at the $Al_2O_3/Hf_{0.5}Zr_{0.5}O_2$ interface. This work provides a method of improving memory window for the application of FeFET to multi-bit and low-power memory cells.